\documentclass[traditabstract]{aa}
\usepackage{graphicx}
\usepackage{txfonts}
\usepackage{natbib}

\bibpunct{(}{)}{;}{a}{}{,} 

 \newcommand{\mics}{$\mu$m~}

\def\tex {\ifmmode{{T}_{\rm ex}}\else{$T_{\rm ex}$}\fi}
\def\tmb {\ifmmode{{T}_{\rm mb}}\else{$T_{\rm mb}$}\fi}
\def\ci     {\ifmmode{{\rm C}{\rm \small I}}\else{C\ts {\scriptsize I}}\fi}
\def\hi     {\ifmmode{{\rm H}{\rm \small I}}\else{H\ts {\scriptsize I}}\fi}
\def\hh     {\ifmmode{{\rm H}_2}\else{H$_2$}\fi}

\def\ts     {\thinspace}
\def\kms    {\ifmmode{{\rm \ts km\ts s}^{-1}}\else{\ts km\ts s$^{-1}$}\fi}
\def\msol   {\ifmmode{{\rm M}_{\odot}}\else{M$_{\odot}$}\fi}
\def\lsol   {\ifmmode{{\rm L}_{\odot}}\else{L$_{\odot}$}\fi}
\def\zsol   {\ifmmode{{\rm Z}_{\odot}}\else{Z$_{\odot}$}\fi}

\begin{document}

\title{Cold gas in group-dominant elliptical galaxies
\thanks{Based on observations carried out with the IRAM 30m telescope.
IRAM is supported by INSU/CNRS (France), MPG (Germany), and IGN (Spain)}}

\author{E. O'Sullivan \inst{1}
\and
F. Combes \inst{2}
\and
S. Hamer \inst{2}
\and
P. Salom\'e \inst{2}
\and
A. Babul \inst{3}
\and
S. Raychaudhury \inst{4,5}
           }
\offprints{E. O'Sullivan}
\institute{Harvard-Smithsonian Center for Astrophysics, 60 Garden Street, MS-50
Cambridge, MA 02138, USA
\email{eosullivan@cfa.harvard.edu}
 \and
Observatoire de Paris, LERMA (CNRS:UMR8112), 61 Av. de l'Observatoire, F-75014, Paris, France
\and
Department of Physics and Astronomy, University of Victoria, Victoria, BC, V8W 2Y2, Canada
 \and
Department of Physics, Presidency University, 86/1 College Street, 700073 Kolkata, India
\and
School of Physics and Astronomy, University of Birmingham, Edgbaston, Birmingham, B15 2TT, UK
              }

\date{Received  2014/ Accepted  2014}

\titlerunning{CO in group-dominant ellipticals}
\authorrunning{E. O'Sullivan et al.}

\abstract{We present IRAM 30m telescope observations of the CO(1-0) and
  (2-1) lines in a sample of 11 group-dominant elliptical galaxies selected
  from the CLoGS nearby groups sample. Our observations confirm the
  presence of molecular gas in 4 of the 11 galaxies at $>$4$\sigma$
  significance, and combining these with data from the literature we find a
  detection rate of 43$\pm$14\%, comparable to the detection rate for
  nearby radio galaxies, suggesting that group-dominant ellipticals may be
  more likely to contain molecular gas than their non-central counterparts.
  Those group-dominant galaxies which are detected typically contain
  $\sim$2$\times$10$^8$~M$_\odot$ of molecular gas, and although most have
  low star formation rates ($<$1~M$_\odot$~yr$^{-1}$) they have short
  depletion times, indicating that the gas must be replenished on
  timescales $\sim$10$^8$~yr. Almost all of the galaxies contain active
  nuclei, and we note while the data suggest that CO may be more common in
  the most radio-loud galaxies, the mass of molecular gas required to power
  the active nuclei through accretion is small compared to the masses
  observed. We consider possible origin mechanisms for the gas, through
  cooling of stellar ejecta within the galaxies, group-scale cooling flows,
  and gas-rich mergers, and find probable examples of each type within our
  sample, confirming that a variety of processes act to drive the build up
  of molecular gas in group-dominant ellipticals.

\keywords{Galaxies: evolution --- Galaxies: general --- Galaxies: groups: general --- Galaxies: kinematics and dynamics --- Galaxies: star formation}
}
\maketitle


\section{Introduction}

Despite the traditional picture of elliptical galaxies as red and dead,
observations over the past decade have shown that a significant fraction of
the population contain quantities of molecular and atomic gas which can
fuel star formation or AGN activity. A molecular gas survey of the
ATLAS$^{3D}$ sample, consisting of ellipticals within 42 Mpc, detects CO in
56/259 galaxies \citep{Youngetal11}, with the origin of the gas likely a
mixture of acquisition during mergers and cooling of material ejected from
the stellar population \citep{Davisetal11}. Cool gas is also associated
with the central dominant ellipticals of galaxy clusters
\citep[e.g.,][]{Edgeetal01,SalomeCombes03} and is generally believed to be
the product of cooling from the hot, X-ray emitting intracluster medium
(ICM).  Ongoing star formation is observed in a number of central
  dominant ellipticals in cool-core galaxy clusters
  \citep{Bildfelletal08,Pipinoetal09,Donahueetal11,McDonaldetal11},
  probably fuelled by material cooled from the ICM. Whereas cold gas in
individual ellipticals typically forms rings or disks, the material around
cluster-dominant galaxies is often found in extended filaments
\citep[e.g.,][]{Salomeetal06,McDonaldetal11,Limetal12}.

Galaxy groups, representing the intermediate mass range between individual
galaxies and massive clusters, offer an interesting avenue for
investigation. Many possess a hot ($\sim$10$^7$~K) intra-group medium (IGM)
and among the (statistically incomplete) ensemble of nearby groups where a
hot IGM has been confirmed via \textit{Chandra} X-ray observations, $>$80\%
have central cool cores \citep{Dongetal10}. The low velocity dispersions
typical in groups make galaxy mergers more common than in clusters and
reduce the efficiency of ram-pressure stripping, suggesting that a greater
fraction of infalling cold-gas-rich galaxies are likely to reach the
group core intact. Since both mergers and cooling flows appear to be common
in groups, a significant fraction of the giant ellipticals commonly
  found at or near the centres of relaxed groups (hereafter referred to as
  group-dominant ellipticals) should contain cold gas.

The presence of molecular gas in group-dominant ellipticals is important to
our understanding of feedback from active galactic nuclei (AGN). As in
cool-core clusters, radiative cooling of the IGM is thought to be regulated
by FR-I radio sources in the group-dominant ellipticals. In some cases cold
gas is already known to be associated with AGN outbursts in groups, for
example in NGC 315 \citep{OcanaFlaqueretal10} or the well-known 100 pc disk
of H\textsc{i} gas and dust in NGC 4261 \citep{JaffeMcNamara94}.
Identification of reservoirs of molecular gas in a sample of group-central
ellipticals has the potential to shed light on the origin of the gas,
provide estimates of the mass available to fuel AGN, and help clarify the
mechanisms of IGM cooling and AGN feedback

We have therefore undertaken a programme of observations using the IRAM 30m
telescope with the goal of identifying the presence of CO in a sample of
group-dominant ellipticals, determining what fraction of these systems
contain molecular gas, and examining the relationship between cold gas, AGN, and group properties. In this paper we present our initial results.

To compute distances, we adopt a standard flat cosmological model, with
$\Lambda$ = 0.75 and a Hubble constant of 70\,km\,s$^{-1}$\,Mpc$^{-1}$.
Our sample and observations are described in Sections~\ref{sample} and
\ref{obs} respectively. Results are presented in Sect. \ref{res} and
discussed in Sect. \ref{disc}.

\section{Sample}
\label{sample}

%
\begin{center}
\begin{table*}
      \caption[]{Basic data for the group-dominant galaxies} 
\label{tab:basic}
\centering
\begin{tabular}{lrrrrrrrrrrr}
\hline\hline
Galaxy  & z  & D$_{L}$ & (1-0) Beam & log L$_{\rm B}^a$  & log D$_{25}^a$ & log M$_{\rm dust}^c$ & log M(HI)$^b$ 
& log L$_{\rm FIR}^c$  & log M$_{*}^d$  & M(H$_2$)$^e$ &  F(1.4GHz)$^f$ \\
        &    &  [Mpc]    & [kpc]     & [\lsol]      &[kpc]       &[\lsol]      &[\lsol]
 &[\lsol]     &[\msol]  & [10$^8$\msol]&  [mJy] \\[+1mm]
\hline\\[-3mm]
\multicolumn{11}{l}{\textit{Newly observed systems}}\\
NGC 193  & 0.014723  & 63.8 & 6.99   & 10.24  &  1.58  &  --     &  --     & --   &10.92 & $<$1.35 & 1710 \\
NGC 677  & 0.017012  & 73.9 & 8.08   & 10.52  &  1.58  &   --    &  --     & --   &10.91 & $<$2.25 & 21\\
NGC 777  & 0.016728  & 72.6 & 7.94   & 11.03  &  1.78  & 3.60   &  --     & 7.91 &11.32 & $<$2.08 & 7\\
NGC 940  & 0.017075  & 74.2 & 8.11   & 10.81  &  1.53  & 6.71  &   --    &10.06&10.94 & 61.0 & 8 \\
NGC 1060 & 0.017312  & 75.2 & 8.22   & 11.09  &  1.71  & 6.49  &   --    &10.51&11.42 & $<$0.76 &9\\
NGC 1167$^i$ & 0.016495  & 71.6 & 7.83   & 10.88  &  1.72  & 7.23 &10.01$^g$& 9.59&11.23 & 3.3 &1840\\
NGC 1587 & 0.012322  & 53.3 & 5.85   & 10.64  &  1.48  &  --     & 9.40  & --   &11.01 & 2.3 &131\\
NGC 2768$^i$ & 0.004580  & 19.7 & 2.18   & 10.28  &  1.50  & 4.36 &7.81$^h$ & 8.66&10.76 & 0.18 &15\\
NGC 5846 & 0.005717  & 24.6 & 2.72   & 10.71  &  1.48  & 4.60  &  8.65 & 7.83&10.53 & $<$0.20 &21\\
NGC 5982 & 0.010064  & 43.5 & 4.78   & 10.75  &  1.60  & 5.48  &  --     & 7.87&10.91 & $<$0.25 & $<$2.5 \\
NGC 7619 & 0.012549  & 54.3 & 5.96   & 10.93  &  1.62  &  --     &  --     & --  &11.21 & $<$0.33 &20\\
\hline
\multicolumn{11}{l}{\textit{CO detected by previous studies}}\\
NGC 315  & 0.016485  & 71.6 & 7.83   & 10.87  & 1.79 & 5.87    &  --     & 9.56 & 11.49  & 0.74 & 6630 \\
NGC 524  & 0.008016  & 34.6 & 3.81   & 10.57  & 1.55 & 6.06    &  --     & 9.39 &  11.17 & 1.9 & 3  \\
NGC 3665 & 0.006901  & 29.7 &  3.28  & 10.18  & 1.54 & 6.68    &  --     & 9.74 &  10.72 & 6.0 & 133 \\
NGC 5044 & 0.009280  & 40.1 &  4.41  & 10.61  & 1.67 & 4.57    &  --     & 8.66 &  11.08 & 0.5 & 36 \\
NGC 5127 & 0.016218  & 70.4 &  7.70  & 10.25  & 1.64 &  --       &  --     &  --     &  10.84 & 0.77 & 1980 \\
NGC 7252 & 0.015984  & 69.4 &  7.59  & 10.55  & 1.57 & 7.02    &  9.61 & 10.67& 10.92 & 58.0 &  25 \\
\hline
\multicolumn{11}{l}{\textit{Upper limits on CO from previous studies}}\\
NGC 1407 & 0.005934   & 25.5 & 2.82   & 10.56  & 1.67 & 5.46   &  --   & 8.49  & 11.09 & $<$0.34 & 89 \\ 
NGC 3613 & 0.006841   & 29.5 & 3.25   & 10.17  & 1.48 & --     &  --   & --    & 10.60 & $<$0.46 & $<$1.3 \\
NGC 4261$^i$ & 0.007378   & 31.8 & 3.51   & 10.41  & 1.60 & 5.20   &  --   & 8.34  & 11.07 & $<$0.48 & 19500 \\
NGC 4697 & 0.004140   & 17.8 & 1.97   & 10.36  & 1.57 & 5.40   &  --   & 8.63  & 10.93 & $<$0.07 & 0.4 \\
NGC 5322 & 0.005937   & 25.6 & 2.82   & 10.33  & 1.62 & 7.53   &  --   & 9.53  & 10.93 & $<$0.58 & 78 \\
NGC 5353 & 0.007755   & 33.4 & 3.69   & 10.18  & 1.35 & 6.20   &  --   & 9.19  & 10.99 & $<$1.32 & 41 \\
\hline
\end{tabular}
\begin{list}{}{}
\item[$^{a}$] Computed from HYPERLEDA (http://leda.univ-lyon1.fr/) 
\item[$^{b}$] These quantities are from NED (http://nedwww.ipac.caltech.edu/) 
\item[$^{c}$] The derivation of L$_{\rm FIR}$ and dust masses M$_{\rm dust}$ is described in Sec. \ref{CO-FIR}.
\item[$^{d}$] Stellar masses were obtained through SED fitting with SDSS and/or 2MASS fluxes
\item[$^{e}$] Molecular gas masses and upper limits either derived from our
  data (see Sections~\ref{COdet} and \ref{CO-mass}) or drawn from previous
  observations
  \citep{Duprazetal90,Sageetal07,OcanaFlaqueretal10,Youngetal11,Davidetal14}.
\item[$^{f}$] 1.4~GHz continuum fluxes are drawn from the NRAO VLA Sky Survey \citep[NVSS,][]{Condonetal98,Condonetal02}\, except in the cases of NGC~940 and NGC~4697, where we extrapolate from measurements at 2.4~GHz \citep{DresselCondon78} and 8.5~GHz \citep{Wrobeletal08} respectively, assuming a spectral index of 0.8. The two upper limits are taken from \citep{Brownetal11}.
\item[$^{g}$] from \citet{Struveetal10} 
\item[$^{h}$] from \citet{Serraetal12}
\item[$^{i}$] The CO content of these systems is discussed individually in Appendix A.
\end{list}
\end{table*}
\end{center}

Representative samples of galaxy groups in the local Universe are
relatively difficult to construct. Since groups contain only a small number
of galaxies, optically-selected group catalogues typically include a
significant fraction of uncollapsed systems and false associations caused
by chance superpositions along the line of sight. X-ray selection,
which confirms the presence of a gravitational potential capable of heating
and retaining an IGM, is in principle more reliable. However, nearby X-ray
samples selected from the ROSAT All-Sky Survey suffer from a bias toward
centrally-concentrated systems \citep{Eckertetal11}, and may not be
representative of the general population, while samples drawn from the
\textit{Chandra} and \textit{XMM-Newton} archives reflect the preference of observers toward luminous and unusual systems.

We draw our groups from the Complete Local-Volume Groups Sample (CLoGS,
O'Sullivan et al., in prep.) which consists of 53 groups in the local
Universe (D$<$80 Mpc). The sample is based on the \citet{Garcia93}
catalogue, which was selected via friends-of-friends and hierarchical
clustering algorithms from the all-sky LEDA galaxy catalogue, complete to
$m_B$=15. The goal of CLoGS is to determine whether these groups contain a
hot IGM via X-ray observations, providing additional information on whether
they are relaxed, virialized systems; to date $\sim$70\% of the sample has
been observed by \textit{Chandra} and/or \textit{XMM-Newton}. The sample is
limited to groups with declination $>$-30\degr\ to ensure coverage by the
Giant Metrewave Radio Telescope (GMRT) and Very Large Array (VLA), so as to
allow identification of active galactic nuclei and star-forming galaxies.
The most $K$-band luminous early-type galaxy associated with the main peak
of the galaxy density distribution is identified as the group-dominant
galaxy.

CLoGS is intended to be a statistically complete survey of groups in the
local volume, and thus provides an ideal sample in which to study the
prevalence and origin of molecular gas in group-dominant galaxies. However,
since this paper describes preliminary observations, the subsample
discussed here is necessarily incomplete. For our initial list of targets,
we selected group-dominant ellipticals that are either detected at 60 or
100 $\mu$m by IRAS (since far-infrared emission suggests the presence of
dust, which is a good indicator that cool gas may also be present), or
which have been observed in the X-ray by \textit{Chandra} or
\textit{XMM-Newton} (since an X-ray luminous IGM, if detected, should
support cooling). This resulted in a list of 12 targets visible from the
IRAM 30m telescope, of which 11 were observed. The basic properties of each
target galaxy are listed in Table \ref{tab:basic}.

\bigskip
\section{Observations}
\label{obs}

The observations were carried out with the IRAM 30m telescope
at Pico Veleta, Spain, in two runs during May and August 2013.
All sources were observed simultaneously in CO(1-0) and
CO(2-1) lines, with the 3mm and 1mm receivers in parallel.

\begin{table*}[!ht]
      \caption[]{Molecular data and stellar mass of the four detected galaxies.
For each object, the first line displays the CO(1-0) and the second line the CO(2-1) results.}
\label{tab:det}
\centering
\begin{tabular}{lcccccl}
\hline\hline
Galaxy & Area &  V & $\Delta$V$^{(1)}$ & T$_{mb}^{(2)}$ & L$^\prime_{\rm CO}$/10$^{8}$& M(H$_2$)$^{(3)}$\\
 &  K km/s &   km/s &   km/s  & mK &   K.km/s.pc$^2$ &  10$^8$\msol\\
\hline
NGC 940  &  19.9$\pm$ 0.7  & 97$\pm$ 8 & 442$\pm$ 15 &42. &13.2 & 61. \\
    &  15.8$\pm$ 0.6  & 129$\pm$ 7  &344$\pm$ 13 & 43. &2.6 & \\
NGC 1167 &  1.18$\pm$ 0.1  & 4$\pm$ 14 &233$\pm$ 47 &4.8 &0.73 & 3.3 \\
    &  1.18$\pm$ 0.2  &  -90$\pm$ 26  &265$\pm$ 60 &4.2 &0.18 & \\
NGC 1587 &  1.43$\pm$ 0.3  & 9$\pm$ 26 &190$\pm$ 44 &7.0 & 0.49& 2.3 \\
    &  0.77$\pm$ 0.3  &  0$\pm$ 50  &205$\pm$ 70 &3.5 &0.1 & \\
NGC 2768 &  0.84$\pm$ 0.3  &-80$\pm$ 60  &340$\pm$ 120 &2.3 & 0.04& 0.18\\
    &  2.5$\pm$ 0.6  &-45$\pm$ 35  &251$\pm$ 58 &9.5 & 0.03& \\
\hline
\end{tabular}
\begin{list}{}{}
\item[]   Results of the Gaussian fits
\item[] $^{(1)}$ FWHM 
\item[] $^{(2)}$ Peak brightness temperature
\item[] $^{(3)}$ obtained with the standard MW conversion ratio
\end{list}
\end{table*}

The broadband EMIR receivers were tuned in single sideband mode, with a
total bandwidth of 4 GHz per polarization. This covers a velocity range of
$\sim$ 10,400 \kms at 2.6mm and $\sim$ 5,200 \kms at 1.3mm.  The
observations were carried out in wobbler switching mode, with reference
positions offset by $2\arcmin$ in azimuth. Several backends were used in
parallel, the WILMA autocorrelator with $2$~MHz channel width, covering
4$\times$4\,GHz, and the Fourier Transform Spectrometer (FTS), covering
8$\times$4\,GHz.

We spent on average two hours on each galaxy, and reached a noise level
between 1 and 3 mK (antenna temperature), smoothed over $30$~km~s$^{-1}$
channels for all sources.  Pointing measurements were carried out every two
hours on continuum sources and the derived pointing accuracy was 3$''$ rms.
The temperature scale is then transformed from antenna temperature $T_{\rm
  A}^*$ to main beam temperature $T_{\rm mb}$, by multiplying by 1.20 at
3mm and 1.56 at 1.3mm.  To convert the signals to fluxes, we use $S/T_{\rm
  mb}$ = 5.0 Jy/K for all bands.  At 2.6mm and 1.3mm, the telescope
half-power beam width is 23$''$ and 12$''$ respectively.  The data were
reduced with the CLASS/GILDAS software, and the spectra were smoothed so
that each line covers about ten channels in the plots.  \footnote{Spectra
  of detections will be available in electronic form at the CDS via
  anonymous ftp to cdsarc.u-strasbg.fr (130.79.128.5) or via
  http://cdsweb.u-strasbg.fr/cgi-bin/qcat?J/A+A/ }

We note that our observations reach sensitivities similar to that of the ATLAS$^{3D}$ sample. \citet{Youngetal11} smoothed their observations over 31~km~s$^{-1}$ channels and reached typical a noise level of 3~mK. While our sample extends to 80~Mpc, ATLAS$^{3D}$ only includes galaxies within 42~Mpc, so our slightly longer integrations lead to very similar limits on detectable gas masses; we find that our observations are typically able to detect molecular gas masses of M(H$_2$)$>$8$\times$10$^7$M$_\odot$, and $>$2.8$\times$10$^7$$_\odot$ for systems within 45~Mpc, whereas the mean limit for the ATLAS$^{3D}$ sample is M(H$_2$)$>$2.3$\times$10$^7$M$_\odot$. The distribution of molecular gas masses and upper limits with distance appears comparable for the two samples. We therefore conclude that our observations can be directly compared with the ATLAS$^{3D}$ sample without introducing strong bias.

\section{Results}
\label{res}

\subsection{CO detections}
\label{COdet}

\begin{figure*}[ht]
\centerline{
\includegraphics[angle=0,width=15cm]{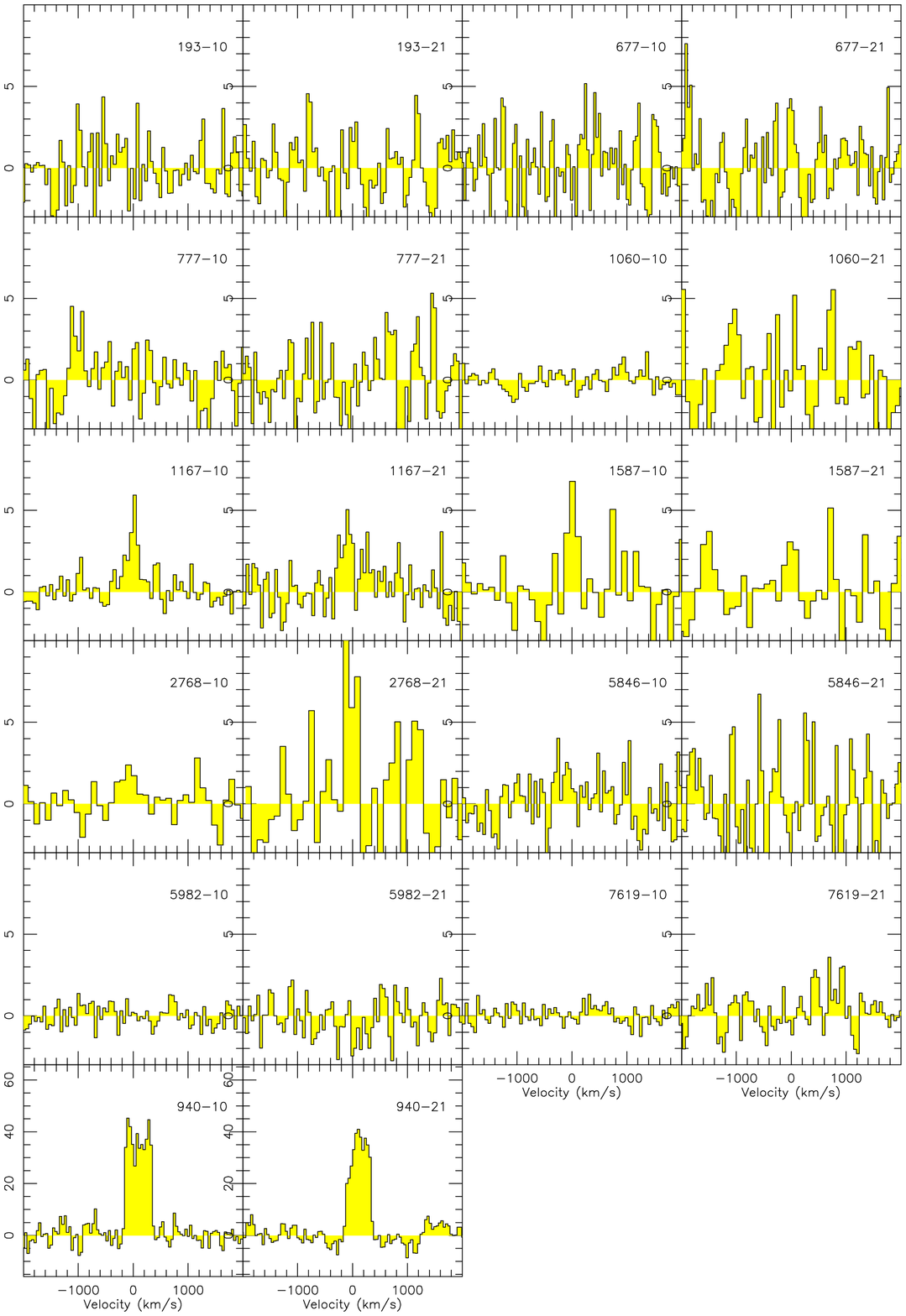}}
\caption{ CO spectra of the observed group-dominant galaxies. In each column,
the left-hand panel is the CO(1-0), and right-hand panel the CO(2-1) lines. The
velocity scale is relative to the redshift displayed in Table \ref{tab:basic}.
The vertical scale is T$_{mb}$ in mK. All spectra have been plotted to the same scale,
except the CO-rich NGC 940 galaxy, at the bottom.}
\label{fig:spectra}
\end{figure*}

Out of the 11 galaxies observed, 4 were detected at $>$4$\sigma$
significance (see Tables~\ref{tab:basic} and \ref{tab:det}). All spectra
are plotted in Figure \ref{fig:spectra} with the same flux scale, with the
exception of NGC~940, since its intensity is about 10 times larger than the
others. A number of other CLoGS group-dominant elliptical galaxies have
already been observed in CO (primarily as part of the ATLAS$^{3D}$ sample,
see Table~\ref{tab:basic}) and we include these systems to increase the
diagnostic power of our subsample.

In total, over the 23 objects for which CO data are available, 10 are
confirmed to contain molecular gas, giving a detection rate of $43\pm14$\%.
The detection rate for early-type galaxies in the ATLAS$^{3D}$ sample is
$22\pm$3\% \citep{Youngetal11}, suggesting that group-dominant galaxies may
be more likely to contain CO, but with our current small sample, the
difference is only significant at the $\sim$1.5$\sigma$ level. Completing
observations of the CLoGS sample will improve the statistical power of our
sample and allow us to determine whether this difference is truly
significant.


\subsection{CO luminosity and \hh\, mass}
\label{CO-mass}

We have computed the molecular mass from the CO(1-0) flux, using M$_{\rm
  H_2} = \alpha$ L$^\prime_{\rm CO}$, with $\alpha=4.6$ M$_\odot$ (K \kms\,
pc$^2$)$^{-1}$, the standard factor for nearby quiescent galaxies like the
Milky Way.  The molecular gas masses are listed in Table~\ref{tab:det} and
the upper-limits in Table~\ref{tab:uplim}. We note that \citet{Youngetal11}
adopted the same value of $\alpha$ when estimating M$_{\rm H_2}$ for the
galaxies observed in the ATLAS$^{3D}$ CO survey.

The average CO luminosity for the four galaxies detected
is L$^\prime_{\rm CO}$ = 3.6$\times$10$^{8}$ K \kms\, pc$^2$, corresponding to an 
average \hh\, mass of  16.7$\times$10$^{8}$ \msol.

Since the beam size of our observations is typically smaller than the
stellar extent of the galaxies, we cannot be certain that we are capturing
the total mass of CO in each system. However, interferometric mapping of CO
has shown that it is typically concentrated in the galaxy core, with a
distribution exponentially declining with radius \citep{YoungScoville91}.
Mapping of 40 ATLAS$^{3D}$ galaxies with detected CO shows that in almost
every case CO is only observed in the central 4~kpc of the galaxy, and only
a handful of galaxies have CO extent $>$2~kpc \citep{Davisetal13}.
Comparing these values to the projected beam sizes in
Table~\ref{tab:basic}, we find that only five of our detections are close
enough for CO to potentially extend beyond the 30m beam. Of these, NGC~2768
has been mapped by the Plateau de Bure Interferometer, which showed that
the 30m mass estimate was accurate \citep{Crockeretal08}, while for
NGC~5044 we have adopted the ALMA interferometric mass estimate
\citep{Davidetal14}. Of the remaining three detected galaxies for which the
30m might underestimate the molecular gas mass, NGC~2768 has been mapped by
CARMA, which found a mass a factor of 1.6 larger than the 30m estimate
\citep{Alataloetal13}. We are therefore confident that for our sample, the
IRAM 30m observations should include the great majority of the molecular
gas, with at most three galaxies having underestimated masses.

\begin{table}
\caption[]{Upper limits on molecular gas mass for galaxies undetected in CO(1-0) and CO(2-1).}
\label{tab:uplim}
\centering
\begin{tabular}{l c c c c }
\hline\hline
\noalign{\smallskip}
Galaxy &Line& rms & L$^\prime_{\rm CO}$/10$^{8}$& M(H$_2$) \\
       &     &  [mK]  & [K \kms\, pc$^2$] &  [10$^8$\msol]\\
\noalign{\smallskip}
\hline
\noalign{\smallskip}
NGC 193  &  CO(1-0)   &  2.1  &  0.29  & 1.35 \\
        &  CO(2-1)     &  2.7  &  0.09  & 0.43 \\
NGC 677  &  CO(1-0)    &  2.6  &  0.49  & 2.25 \\
        &  CO(2-1)     &  2.8  &  0.13  & 0.61 \\
NGC 777  &  CO(1-0)    &  2.5  &  0.45  & 2.08 \\
        &  CO(2-1)     &  2.8  &  0.13  & 0.58 \\
NGC 1060 &  CO(1-0)  &  0.8  &  0.16  & 0.76 \\
        &  CO(2-1)     &  4.3  &  0.21  & 0.96 \\
NGC 5846 &  CO(1-0)  &  2.1  &  0.04  & 0.20 \\
        &  CO(2-1)     &  3.5  &  0.02  & 0.08 \\
NGC 5982 &  CO(1-0)  &  0.8  &  0.06  & 0.25 \\
        &  CO(2-1)     &  1.6  &  0.03  & 0.12 \\
NGC 7619 &  CO(1-0)  &  0.7  &  0.07  & 0.33 \\
        &  CO(2-1)     &  1.5  &  0.04  & 0.17 \\
\noalign{\smallskip}
\hline
\end{tabular}
\begin{list}{}{}
\item[]  The rms are in T$_{mb}$ in  channels of 30 \kms.
\item[] The upper limits in L$^\prime_{\rm CO}$ and M(\hh) are at 3$\sigma$ with an assumed $\Delta$V = 300 \kms.
\end{list}
\end{table}

\subsection{Star Formation}
\label{CO-FIR}

The far infrared luminosities were computed from the 60\mics and 100\mics
fluxes from IRAS, or when absent from the 24, 70 and 160\mics fluxes from
Spitzer/MIPS.  These flux ratios give the dust temperature, assuming a mass
opacity of the dust at frequency $\nu$, $\kappa_\nu\propto\nu^{\beta}$,
where $\beta$ = 1.5.  For galaxies with far infrared detections, the dust
mass has been estimated as
$$
\begin{array}{lcl}
 {\rm M}_d & = & 4.8\times10^{-11}\, {{S_{\nu o}\,D_{\rm Mpc}^{\,2}}
 \over {(1+z)\kappa_{\nu r}\,B_{\nu r}(T_d)}}\ \msol \\
\end{array}
$$
where $S_{\nu o}$ is the observed FIR flux measured in Jy, $D_{\rm
  Mpc}^{\,2}$ the luminosity distance in Mpc, $B_{\nu r}$ the Planck
function at the rest frequency $\nu r = \nu o (1+z)$, and we use a mass
opacity coefficient of $25$~cm$^{2}$~g$^{-1}$ at rest frame 100~\mics
\citep{Hildebrand83,Dunneetal00,Draine03}.

For those galaxies detected in the far infrared, we estimate the star
formation rate SFR by the relation SFR= L$_{\rm FIR}$
/(5.8$\times$10$^9$L$_{\odot}$) compiled by \citet{Kennicutt98}.  The gas
consumption time scale can then be derived as $\tau$ = 5.8 M(\hh)/L$_{\rm
  FIR}$ Gyr, where masses and luminosities are in solar units.  These
values are displayed in Table \ref{tab:SFR}. We note that the depletion
timescales only take account of the detected molecular gas, and that for
H\textsc{i}-rich galaxies such as NGC~1167 they will be an underestimate of
the true gas depletion time. However, most of our targets are
H\textsc{i}-poor, and it is notable that the majority of them have small
depletion time-scales, despite very low specific star formation rates (sSFR
= SFR/M$_*$). Even the galaxies where only upper limits on molecular gas
content are available typically have depletion times $<$10$^9$~yr. This
indicates that, as expected, most of these group-dominant galaxies have
formed the bulk of their stars in earlier star formation episodes.

\begin{figure}[!ht]
\centering
\includegraphics[angle=0,width=8cm]{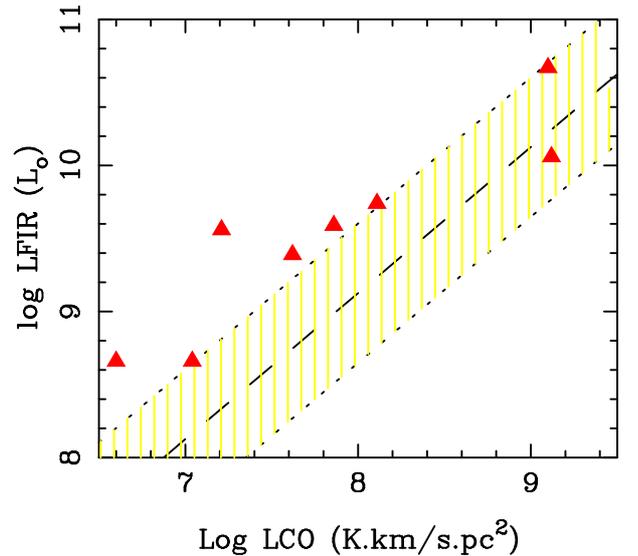}
\caption{Correlation between FIR and CO luminosities, for
the CLoGS sample (filled red triangles).
The dashed line represents a depletion time-scale of 2 Gyr,
assuming a conversion factor
$\alpha$ = 4.6  M$_\odot$ (K \kms\, pc$^2$)$^{-1}$.
The shaded regions represent the main location of nearby
spiral galaxies \protect\citep[from][]{Bigieletal08}, within a factor 3 around
the mean depletion time-scale of 2 Gyr.
}
\label{fig:FIR-CO}
\end{figure}

Figure~\ref{fig:FIR-CO} shows the relation between far infrared and CO
luminosities. The detected objects have a tendency to lie above the normal
location for nearby spiral galaxies, which corresponds to a short depletion
time-scale. Since most of the galaxies contain radio-luminous AGN (see
Section~\ref{AGN}) we might expect some contribution to FIR luminosity from
nuclear activity. However, we see no clear correlation between radio
luminosity and position in Figure~\ref{fig:FIR-CO}. NGC~5044, NGC~3665 and
NGC~524, whose radio luminosities differ by a factor $\sim$30, fall at
similar locations relative to the depletion timescale relation for spirals.
The two galaxies which fall further above the relation, NGC~315 and
NGC~2768, have radio luminosities which differ by a factor $\sim$5800.  We
therefore conclude that the AGN contribution to FIR luminosity has a
minimal effect on the comparison with CO luminosity.

\begin{table}
\caption[]{Star formation rates and depletion times}
\label{tab:SFR}
\centering
\begin{tabular}{l c c c }
\hline\hline
\noalign{\smallskip}
Galaxy &  SFR &  $\tau$ &   sSFR \\
 &  [\msol/yr]  & [Gyr] &  [Gyr$^{-1}$]\\
\noalign{\smallskip}
\hline
\noalign{\smallskip}
NGC 777  &  0.014  &  --    &  6.7E-5   \\
NGC 940  &  2.0    &  3.1   &  2.3E-2   \\
NGC 1060 &  5.6    &  --    &  2.1E-2   \\
NGC 1167 &  0.67   &  0.49  &  3.9E-3   \\
NGC 2768 &  0.079  &  0.23  &  1.4E-3   \\
NGC 5846 &  0.012  &  --    &  3.5E-4   \\
NGC 5982 &  0.013  &  --    &  1.6E-4   \\
\hline
NGC 315  &  0.62   &  0.12  &  2.0E-3   \\
NGC 524  &  0.42   &  0.45  &  2.8E-3   \\
NGC 3665 &  0.95   &  0.63  &  1.8E-2  \\
NGC 5044 &  0.079  &  0.63  &  6.6E-4  \\
NGC 7252 &  8.1    &  0.72  &  9.7E-2   \\
\noalign{\smallskip}
\hline
\end{tabular}
\end{table}

\begin{figure}[!ht]
\centering
\includegraphics[angle=0,width=8cm]{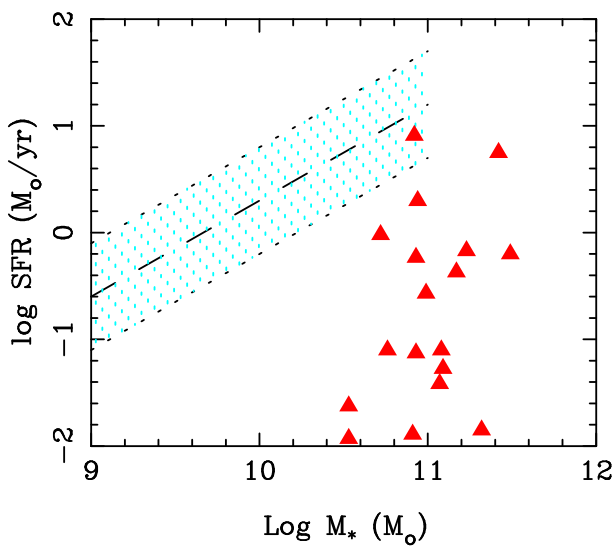}
\caption{ Position of our sample galaxies in the 
SFR-stellar mass diagram. The blue shaded region indicates
the main sequence of star forming galaxies at z=0 
\citep{Wuytsetal11}.
}
\label{fig:MS-SFR}
\end{figure}

Figure \ref{fig:MS-SFR} shows the relation between stellar mass and SFR
(for galaxies detected in the far-infrared). To estimate stellar masses, we
determined galaxy colours from Sloan Digital Sky Survey $ugriz$ and 2MASS
$JHK$ magnitudes, and used these to define the mass-to-light ratio of each
galaxy from the models of \citet{BelldeJong01}.  The main sequence of star
forming galaxies at z=0 is indicated by the blue shaded region
\citep[from][]{Wuytsetal11}.  As expected, the galaxies of our sample
mainly lie in the red sequence of quenched galaxies. However, a handful of
the most rapidly star-forming approaching the Wuyts et al. relation,
suggesting that their specific star formation rates are comparable with
those of massive spirals. This is a surprising result for massive,
group-dominant elliptical galaxies, but may indicate the importance of cold
molecular gas, the fuel for star formation, in the development of these
systems.

\subsection{Group properties}

For those systems which have high-quality X-ray data available, we can
attempt to determine whether the molecular gas content of the
group-dominant galaxy is related to the properties of any hot, X-ray
emitting intra-group medium. The overlap with our molecular gas sample is
imperfect; only 20 of the 23 galaxies for which CO data are available have
high-quality \textit{Chandra} or \textit{XMM} observations
\citep[O'Sullivan et al., in
prep.]{OSullivanPonman04,Nolanetal04,Helsdonetal05,NaginoMatsushita09,Pellegrini11,Borosonetal11}.
One other galaxy, NGC~524, has been observed by \textit{Chandra}, but only
using the ACIS-S detector in 1/8 subarray mode, providing a very limited
field of view unsuitable for examination of the properties of the group.
However \textit{Rosat} data of sufficient quality to examine the gas
content are available \citep{OsmondPonman04}. For these 21 systems we are
able to determine whether hot, X-ray emitting gas is present as an
intra-group medium (IGM), or as a smaller halo associated with the dominant
galaxy.

We follow the scheme of \citet{OsmondPonman04} in differentiating between
group and galaxy emission based on a cutoff in physical scale; systems
whose diffuse X-ray emission extends $<$65~kpc are classed as galaxy halos,
while more extended systems are considered to possess a true IGM. We
further class systems with a significant decline in central temperature as
having a cool core, indicating that radiative cooling has been more
effective than any heating mechanisms over the recent history of the group.
However, we note that this classification is dependent on the quality of
the data available; fainter systems sometimes lack the necessary
signal-to-noise ratio to allow measurement of the radial temperature
profile.We find that hot gas is detected in 20 of the 21 systems for which
data are available, and 14 of these reside in groups with an extended IGM;
the remaining 6 only have small halos of hot gas associated with the
central galaxy.

Of the 10 galaxies in which CO is detected, one (NGC~5127) has yet to be
observed by \textit{XMM} or \textit{Chandra}. Only 4 of the remaining 9
galaxies are in groups with an extended IGM, and of these, only 2 are in
systems with cool cores. Of the five galaxies with CO detections which do
not reside in groups with an extended hot IGM, four possess galaxy-scale
halos of hot gas, leaving only one galaxy detected in CO but with no hot
gas detection. These results confirm that, as in more massive galaxy
clusters, cold molecular gas is formed in a sizeable fraction of
group-dominant ellipticals, and that it survives for significant periods of
time despite the surrounding hot IGM.

\subsection{Nuclear Activity}
\label{AGN}

Of the 23 galaxies with CO data, all but one appear to host radio and/or
X-ray luminous active nuclei. For the 21 galaxies observed by
\textit{Chandra} or \textit{XMM}, we base our identification of X-ray AGN
either on our own analysis (O'Sullivan et al., in prep.) or on reports in
the literature
\citep{Nolanetal04,Pellegrini05,GonzalezMartinetal06,Borosonetal11,Liuetal11}.
Sources are identified as likely AGN if they are coincident with the galaxy
nucleus to within the spatial resolution of the X-ray data. In all but a
handful of cases, the sources are confirmed to have power-law spectra, and
those for which spectra are not available are all identified from
\textit{Chandra} data, whose 0.5$^{\prime\prime}$ spatial resolution makes
confusion with an extended source (e.g., a star-forming region) unlikely.
We find that $\ga$70\% (17/23) of the galaxies host nuclear X-ray sources
(including NGC~524, whose AGN is detected in the limited field-of-view
\textit{Chandra} observation).

When identifying radio AGN, we primarily rely on high-resolution radio data
to determine whether sources are coincident with the galaxy nucleus, and
whether they are extended. The CLoGS sample was designed to match the sky
coverage of the major VLA surveys, so 1.4~GHz continuum imaging from FIRST
\citep{Beckeretal95} and/or NVSS \citep{Condonetal93NVSS} is available for
every galaxy. In general we use FIRST (spatial resolution
5.4$^{\prime\prime}$ FWHM, equivalent to 0.5-2.0~kpc) to determine whether
a source is likely to be an AGN, but draw flux measurements from NVSS,
since its larger restoring beam (45$^{\prime\prime}$ FWHM) makes it less
prone to break up extended sources into separate components. In addition,
we have deeper \textit{Giant Metrewave Radio Telescope} (GMRT) 610~MHz
images (half-power beam-width HPBW=5-6$^{\prime\prime}$) for 13/23
galaxies.

Table~\ref{tab:basic} lists 1.4~GHz continuum fluxes for the galaxies. All
but four of the galaxies are detected in this band.  NGC~4697 has been
imaged with sub-arcsecond resolution at 8.5~GHz by the VLA, confirming the
presence of a weak nuclear source coincident with a \textit{Chandra} X-ray
source \citep{Wrobeletal08}. NGC~940 is detected at low spatial resolution
(HPBW=2.7$^{\prime}$) in a 2.38~GHz Arecibo observation
\citep{DresselCondon78}. In both cases we estimate a 1.4~GHz continuum flux
for the galaxies assuming a spectral index of 0.8. Only an upper limit on
the 1.4~GHz continuum flux is available for NGC~5982, but GMRT 610~MHz
observations \citep{Kolokythasetal14} confirm the presence of an unresolved
source coincident with the galaxy nucleus. Lastly, while it is detected at
1.4~GHz, only low-resolution NVSS data are available for the post-merger
starburst galaxy NGC~7252. Combining all available data, we find that only
only NGC~3613 is undetected at radio frequencies, but that the low spatial
resolution data available for NGC~940 and NGC~7252 does not allow us to
separate a nuclear source from more extended star formation.

As a further test, we estimate the 1.4~GHz continuum flux density expected
from star formation, based on the star formation rates calculated from the
FIR luminosity in Section~\ref{CO-FIR} and the L$_{\rm 1.4GHz}$:SFR
relation of \citet{Bell03}. \citet{Oosterlooetal10} find that $\sim$50\% of
nearby ellipticals (across a range of environments) contain a central
continuum source, but that in galaxies containing H$\textsc{i}$ in their
central regions, the radio flux is often consistent with the star-formation
rate estimated from the FIR. While the radio luminosities of these
star-forming systems identified by Oosterloo are lower than all but the
weakest of our sources (1-31$\times$10$^{19}$~W~Hz$^{-1}$), it is possible
that in some cases we could mistake star formation emission for an AGN. Of
the 17 galaxies in our sample with FIR data, 12 have radio luminosities
more than a factor of 5 greater than that expected from star formation, and
are therefore clearly AGN-dominated. Of the remaining five, NGC~1060 and
NGC~4697 have radio luminosities lower than expected from their SFR, but
radio imaging confirms them both as AGN. The radio emission in NGC~1060
arises from small-scale AGN jets \citep{Kolokythasetal14}, while in
NGC~4697 the radio emission is clearly associated with the X-ray AGN
\citep{Wrobeletal08}. The third galaxy, NGC~524, has been shown to possess
a central milliarcsecond-scale radio point \citep{Filhoetal04} but
\citet{Oosterlooetal10} argue that a fraction of its radio emission arises
from star formation. In the last two galaxies, NGC~7252 and NGC~940, we
cannot confirm the presence of a radio-luminous AGN, and it is probable
that a significant fraction of the radio emission arises from star
formation. However, both galaxies have X-ray AGN and it seems possible that
at least part of their radio luminosity arises from nuclear activity.

We therefore find that $\ga$87\% of the galaxies ($\ga$20/23) host radio
AGN, and in all $\ga$95\% host AGN of some sort. Only NGC~3613 lacks a
detected AGN, and this may be explained by the fact that it has not been
observed with the current generation of X-ray telescopes.

\begin{figure}
\includegraphics[width=\columnwidth,bb=20 215 570 750]{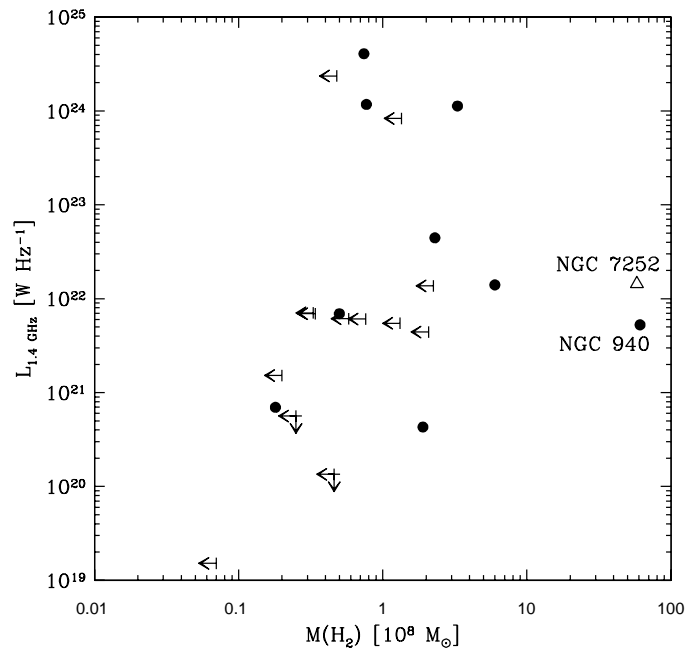}
\caption{\label{fig:Lrad} 1.4~GHz radio continuum luminosity plotted against molecular gas mass. AGN are marked with circles or arrows indicating 3$\sigma$ upper limits in molecular gas mass, or 2$\sigma$ limits in 1.4~GHz luminosity. The open triangle indicates the starburst galaxy NGC~7252. The galaxy with the greatest molecular gas mass among our new observations, NGC~940, is also labelled.}
\end{figure}

Most of the detected AGN are low-luminosity systems; the 1.4~GHz radio
continuum luminosities range from 10$^{19}$-10$^{24}$~W~Hz$^{-1}$, but only 8
galaxies host AGN with L$_{1.4~GHz}>10^{22}$~W~Hz$^{-1}$. 
Figure~\ref{fig:Lrad} shows a comparison of molecular gas mass and 1.4~GHz
radio luminosity for the galaxies in our sample. Although it is unclear
whether there is any direct correlation between gas mass and radio power,
there is a suggestion that a higher fraction of the most radio luminous
galaxies are detected in CO. Excluding NGC~7252, 5 of the 8 galaxies
($\sim$63$\pm$35\%) with L$_{1.4~GHz}>10^{22}$~W~Hz$^{-1}$ are confirmed to
contain CO. The sample is too small for this to be a
  statistically significant result, but we note that altering the radio
  luminosity limit to 10$^{21.5}$ or 10$^{22.5}$~W~Hz$^{-1}$ does not
  change the result; in each case, a larger fraction of radio loud galaxies
  are detected in CO than are radio quiet galaxies. The detection rates are
similar to that found for the TANGO I sample of radio galaxies selected to
have L$_{1.4~GHz}>10^{22.5}$~W~Hz$^{-1}$
\citep[$\sim$40-60\%,][]{OcanaFlaqueretal10}. The CO detection rate among
our less radio luminous systems is only $\sim$29$\pm$14\% (4/14),
comparable to that seen among ellipticals in the ATLAS$^{3D}$ sample
\citep[$\sim$20\%,][]{Youngetal11}.

\section{Summary and discussion}
\label{disc}

We have presented our CO survey in 11 group-dominant elliptical galaxies,
observed with the IRAM-30m telescope. Four galaxies were detected at
$>$4$\sigma$ significance, and when we include data from previous
observations we find a detection rate of 43$\pm$14\%. This result suggests
that group-dominant ellipticals are more likely to contain molecular gas
than the general population of early-type galaxies, but further
observations are needed to improve the statistical power of the sample
before it can be confirmed.

\begin{figure}
\includegraphics[height=\columnwidth,angle=-90]{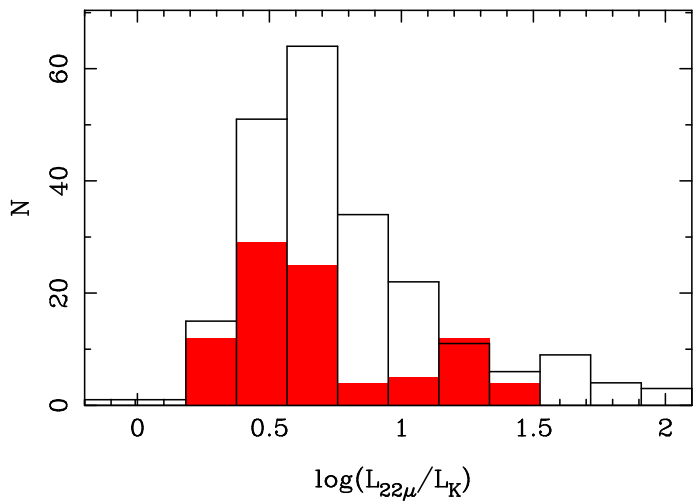}
\caption{\label{fig:L22LK} Histograms of the ratio of 22$\mu$m to $K$-band luminosity (in solar units) for the ATLAS$^{3D}$ sample (black) and our group-dominant galaxies (red). For clarity, the histogram for our sample has been scaled up by a factor of 4.}
\end{figure}

While CLoGS is designed to provide a statistically complete, representative
sample, it is possible that the subset of group-dominant galaxies for which
CO observations are currently available may be biased.  For example, we
selected targets for our CO observing program partly on the basis of FIR
detection, since FIR-luminous galaxies are more likely to be CO-rich. Seven
of our eleven targets are detected in the FIR, as are ten of the twelve
galaxies covered by literature studies, though these were not selected to
be FIR-bright. To determine whether the subsample studied in this paper is
biased, in Figure~\ref{fig:L22LK} we compare the distribution of
22$\mu$m/$K$-band luminosities in our galaxies with that of the
ATLAS$^{3D}$ sample. This provides an indicator of the degree of FIR
emission normalized for differences in galaxy stellar mass.
Figure~\ref{fig:L22LK} suggests that while there is a slight trend toward
higher 22$\mu$m luminosities in the group-dominant galaxies, our sample
covers a similar range to ATLAS$^{3D}$ and is probably not seriously
biased.

Averaging over the 10 galaxies in the sample with measured CO masses, the
mean molecular gas mass is found to be 1.2$\times$10$^{9}$ \msol. However,
the high gas masses in NGC~940 and NGC~7252 have a significant impact on
the mean. If we exclude these galaxies we find a mean molecular gas mass of
2.0$\times$10$^{8}$ \msol. Examining the FIR luminosity and SFR of the
galaxies (and excluding the exceptionally H\textsc{i}-rich system NGC~1167)
we find that most have low star formation rates ($<$1\msol\,yr$^{-1}$) and
short gas depletion timescales. This suggests either that the gas is
replenished on fairly short timescales ($\sim$10$^8$~yr) or that we are
catching the galaxies during a short period of weak star formation which
will soon cease when the supply of cool gas is exhausted. The fraction of
galaxies in which we see short depletion times suggests that $\sim$30\% of
group-dominant galaxies are undergoing a period of star formation at any
one time, suggesting that sufficient cold gas builds up to fuel star
formation approximately three times per Gyr. In either case, our results
indicate that the cold gas reservoirs of the galaxies are replenished on
timescales of a few 10$^8$~yr.

While almost all the group-dominant galaxies contain AGN regardless of
whether they contain molecular gas, the most radio luminous AGN appear to
be more likely to reside in galaxies with larger masses of molecular gas.
Previous studies have shown that the fraction of radio galaxies which
contain molecular gas is higher than that for the general population of
elliptical galaxies, suggesting that molecular gas contributes to the
fuelling of the AGN. However, the accretion rates required to power the
jets of even the most radio luminous galaxies in our sample are quite
modest. Estimates of the total energy of AGN outbursts are available for
NGC~193, NGC~315, NGC~4261 and NGC~5044
\citep{Crostonetal08,Davidetal09,OSullivanetal11b,OSullivanetal11c,Bogdanetal14}
and range from $\sim$10$^{57}$ to $\sim$2$\times$10$^{59}$~erg.  Assuming
an efficiency of 0.01 \citep[c.f.,][]{Babuletal13}, these imply accretion
rates of $\sim$0.5-15$\times$10$^{-4}$~M$_\odot$~yr$^{-1}$, with the total
masses of gas accreted over the outbursts being
$\sim$0.1-20$\times$10$^{6}$~\msol.  While NGC~4261 and NGC~193 are
undetected in CO, both galaxies could contain masses of molecular gas
sufficient to fuel a major outburst without violating the measured upper
limits. Overall, our results are consistent with a scenario in which AGN
outbursts are powered by accretion of relatively small quantities of cold
gas, perhaps in the form of individual clouds, with galaxies containing a
greater mass of cold material being more likely to be observed during the
accretion of a cloud.

For those CLoGS groups confirmed to possess an extended hot intra-group
medium with a cool core, the question arises of whether radiative cooling
from the IGM could provide enough molecular gas to fuel nuclear activity in
the central elliptical. \citet{Davidetal14} examine this question for
NGC~5044 and conclude that as long as more than $\sim$5\% of the gas
cooling in the central 10~kpc of the IGM avoids being reheated, it will
exceed the contribution from stellar mass loss and can support the observed
AGN luminosity. Applying their approach to NGC~315, we estimate the
classical mass deposition rate from radiative cooling of the IGM in the
central 10~kpc to be $\sim$0.6~M$_\odot$~yr$^{-1}$, roughly one tenth that
observed in NGC~5044. We estimate a stellar mass loss rate of
0.17~M$_\odot$~yr$^{-1}$, based on a fit to the 2MASS $K$-band light
profile, adopting a specific stellar mass loss rate of
$\dot{M}_*$/$M_*$=5.4$\times$10$^{-20}$~s$^{-1}$
\citep{RenziniBuzzoni86,Mathews89} and a $K$-band mass-to-light ratio of
0.8~M$_\odot$/$L_{K\odot}$ \citep{Humphreyetal06}. This suggests a finer
balance between stellar mass loss and IGM cooling than in NGC~5044,
requiring that $>$30\% of cooling IGM gas avoid being reheated if it is to
dominate the fuelling of the AGN. However, both stellar mass loss and IGM
cooling rates greatly exceed the likely rate of molecular gas accretion
required to power the AGN.

Three possible origins for cold molecular gas in elliptical galaxies have
been suggested; acquisition through gas-rich merger, cooling from gas
ejected by the stellar population, and cooling from a surrounding IGM.  Our
sample includes the dominant ellipticals from groups with a range of
properties, from loose groups with no detected hot IGM to X-ray luminous
systems with cool cores, and we therefore expected to see evidence of all
three mechanisms in our data. \citet{Davisetal11} argue that gas produced
by the stellar population should form a kiloparsec-scale rotating disk
aligned with the stars, while gas brought into the system through mergers
is likely to be misaligned or to form multiple tails, rings or disks.
Examples of both cases may be present in our sample, for example the
nuclear disk in NGC~4261, and the post-merger galaxy NGC~7252. By analogy
with clusters \citep[e.g.,][]{OSullivanetal12}, we might expect material
cooled from an IGM to form filamentary structures \citep[as in
NGC~5044,][]{Davidetal11}, with molecular gas perhaps forming a rotating
disk in the galaxy core. Of the 14 systems in our sample in which a hot IGM
is observed, 5 are detected in CO, suggesting that IGM cooling may be a
fairly effective production mechanism.  Higher resolution mapping of CO,
and perhaps imaging of other warm and cool gas phases (e.g., H$\alpha$,
H\textsc{i}) would be required to determine whether the gas follows the
morphology expected for IGM cooling.

\begin{acknowledgements}
  We thank M.~Brown for providing 1.4~GHz flux limits for two galaxies, and
  the anonymous referee for comments which have improved the paper.  The
  IRAM staff is gratefully acknowledged for their help in the data
  acquisition.  F.C. and S.H. acknowledge the European Research Council for
  the Advanced Grant Program Number 267399-Momentum.  E.O'S. acknowledges
  support from the National Aeronautics and Space Administration through
  Chandra Award Number AR3-14014X issued by the Chandra X-ray Observatory
  Center, and A.B. from NSERC Canada through the Discovery Grant Program.
  A.B. would like to thank the LERMA group of l'Observatoire de Paris and
  the IAP for hosting his visits. This work made use of the NASA/IPAC
  Extragalactic Database (NED), and of the HyperLeda database.
\end{acknowledgements}

\bibliographystyle{aa}
\bibliography{../paper}

\appendix
\section{Comments on Individual objects}
\label{indiv}

NGC~1167 has a very large HI disk \citep{Struveetal10}, and a large dust
mass. The molecular gas derived from the detected CO emission is however
modest.  Given the usual exponential radial distribution of CO emission in
galaxies \citep[e.g.,][]{YoungScoville91}, it is not likely that we are
missing some CO emission outside of the 7.8kpc beam. The galaxy is currently active, hosting the compact steep spectrum radio source B2~0258+35, and shows signs of previous AGN outbursts in the forms of large-scale ($\sim$10$^{\prime}$ / 200~kpc) low surface brightness lobes \citep{Shulevskietal12}. 

NGC~2768 was already detected in the CO survey of SAURON galaxies
\citep{Combesetal07} and mapped with Plateau de Bure \citep{Crockeretal08}.
The CO maps show that the molecular component lies nearly perpendicular to
the optical major axis of the galaxy. This is obviously gas which has been
recently accreted in a polar ring or disk, and could well come from some
cooling filament.

NGC~4261 was observed with the James Clerk Maxwell Telescope (JCMT) by \citet{JaffeMcNamara94} who reported the detection of CO(2-1) in absorption, located in a nuclear disk, with a peak absorption depth of 14~mK observed against a continuum level of 20mK in the 1.3~mm band. However, later IRAM 30m observations \citep{Combesetal07,Youngetal11} find no indication of CO in either absorption or emission. After re-reducing the IRAM observations, we find a continuum level of 40~mK in the 2.6~mm band and 14~mK in the 1.2~mm band, leading us to expect a continuum level in the JCMT data of 3~mK at 1.3~mm, in strong conflict with the JCMT results. We therefore adopt the IRAM 30m upper limits on molecular gas content for NGC~4261.

\end{document}